\documentclass[twocolumn,showpacs,superscriptaddress,preprintnumbers,amsmath,amssymb]{revtex4}

\usepackage{graphicx}
\usepackage{dcolumn}
\usepackage{bm}
\usepackage[perpage]{footmisc}
\usepackage{natbib}
\usepackage{float}
\usepackage{verbatim}
\usepackage{epstopdf}

\begin{document}



\title{Quenching Factor for Low Energy Nuclear Recoils in a Plastic Scintillator}%
\author{L.~Reichhart\footnote{corresponding author: l.reichhart@sms.ed.ac.uk}} \affiliation{School of Physics \& Astronomy, University of Edinburgh, UK}
\author{D.Yu.~Akimov}\affiliation{Institute for Theoretical and Experimental Physics, Moscow, Russia}
\author{H.M.~Ara\'{u}jo}\affiliation{Blackett Laboratory, Imperial College London, UK}
\author{E.J.~Barnes}\affiliation{School of Physics \& Astronomy, University of Edinburgh, UK}
\author{V.A.~Belov} \affiliation{Institute for Theoretical and Experimental Physics, Moscow, Russia}
\author{A.A.~Burenkov} \affiliation{Institute for Theoretical and Experimental Physics, Moscow, Russia}
\author{V.~Chepel}\affiliation{LIP--Coimbra \& Department of Physics of the University of Coimbra, Portugal}
\author{A.~Currie}\affiliation{Blackett Laboratory, Imperial College London, UK}
\author{L.~DeViveiros}\affiliation{LIP--Coimbra \& Department of Physics of the University of Coimbra, Portugal}
\author{B.~Edwards}\affiliation{Particle Physics Department, STFC Rutherford Appleton Laboratory, Chilton, UK}
\author{V.~Francis}\affiliation{Particle Physics Department, STFC Rutherford Appleton Laboratory, Chilton, UK}
\author{C.~Ghag} \affiliation{School of Physics \& Astronomy, University of Edinburgh, UK}
\author{A.~Hollingsworth} \affiliation{School of Physics \& Astronomy, University of Edinburgh, UK}
\author{M.~Horn}  \affiliation{Blackett Laboratory, Imperial College London, UK}
\author{G.E.~Kalmus} \affiliation{Particle Physics Department, STFC Rutherford Appleton Laboratory, Chilton, UK}
\author{A.S.~Kobyakin} \affiliation{Institute for Theoretical and Experimental Physics, Moscow, Russia}
\author{A.G.~Kovalenko} \affiliation{Institute for Theoretical and Experimental Physics, Moscow, Russia}
\author{V.N.~Lebedenko}  \affiliation{Blackett Laboratory, Imperial College London, UK}
\author{A.~Lindote} \affiliation{LIP--Coimbra \& Department of Physics of the University of Coimbra, Portugal}
\affiliation{Particle Physics Department, STFC Rutherford Appleton Laboratory, Chilton, UK}
\author{M.I.~Lopes} \affiliation{LIP--Coimbra \& Department of Physics of the University of Coimbra, Portugal}
\author{R.~L\"{u}scher} \affiliation{Particle Physics Department, STFC Rutherford Appleton Laboratory, Chilton, UK}
\author{P.~Majewski} \affiliation{Particle Physics Department, STFC Rutherford Appleton Laboratory, Chilton, UK}
\author{A.St\,J.~Murphy} \affiliation{School of Physics \& Astronomy, University of Edinburgh, UK}
\author{F.~Neves} \affiliation{LIP--Coimbra \& Department of Physics of the University of Coimbra, Portugal}
\affiliation{Blackett Laboratory, Imperial College London, UK}
\author{S.M.~Paling}\affiliation{Particle Physics Department, STFC Rutherford Appleton Laboratory, Chilton, UK}
\author{J.~Pinto da Cunha} \affiliation{LIP--Coimbra \& Department of Physics of the University of Coimbra, Portugal}
\author{R.~Preece} \affiliation{Particle Physics Department, STFC Rutherford Appleton Laboratory, Chilton, UK}
\author{J.J.~Quenby}  \affiliation{Blackett Laboratory, Imperial College London, UK}
\author{P.R.~Scovell} \affiliation{School of Physics \& Astronomy, University of Edinburgh, UK}
\author{C.~Silva} \affiliation{LIP--Coimbra \& Department of Physics of the University of Coimbra, Portugal}
\author{V.N.~Solovov} \affiliation{LIP--Coimbra \& Department of Physics of the University of Coimbra, Portugal}
\author{N.J.T.~Smith} \affiliation{Particle Physics Department, STFC Rutherford Appleton Laboratory, Chilton, UK}
\author{P.F.~Smith} \affiliation{Particle Physics Department, STFC Rutherford Appleton Laboratory, Chilton, UK}
\author{V.N.~Stekhanov} \affiliation{Institute for Theoretical and Experimental Physics, Moscow, Russia}
\author{T.J.~Sumner} \affiliation{Blackett Laboratory, Imperial College London, UK}
\author{C.~Thorne} \affiliation{Blackett Laboratory, Imperial College London, UK}
\author{R.J.~Walker} \affiliation{Blackett Laboratory, Imperial College London, UK}
\date{\today}

\begin{abstract}
\noindent 
Plastic scintillators are widely used in industry, medicine and scientific research, including nuclear and particle physics.  Although one of their most common applications is in neutron detection, experimental data on their response to low-energy nuclear recoils are scarce. Here, the relative scintillation efficiency for neutron-induced nuclear recoils in a polystyrene-based plastic scintillator (UPS--923A) is presented, exploring recoil energies between 125~keV and 850~keV. Monte Carlo simulations, incorporating light collection efficiency and energy resolution effects, are used to generate neutron scattering spectra which are matched to observed distributions of scintillation signals to parameterise the energy-dependent quenching factor. At energies above 300 keV the dependence is reasonably described using the semi-empirical formulation of Birks and a $kB$ factor of (0.014$\pm$0.002)~g$\,$MeV$^{-1}$cm$^{-2}$ has been determined. Below that energy the measured quenching factor falls more steeply than predicted by the Birks formalism.
\end{abstract}

\pacs{32.50.+d; 78.70.Ps; 29.40.Mc; 28.20.Cz}
\keywords{plastic scintillator, quenching factor, nuclear recoils, ZEPLIN-III}

\maketitle

\section{Introduction}

\noindent The response of organic scintillators to particle interactions in terms of the dependence on material, incident particle type and incident particle energy were first discussed by Birks~\cite{Birks:1951p2813, Birks:1964}. In general, it is found that the response arising from nuclear recoils (such as when irradiated by neutrons) is significantly diminished in comparison to the light output obtained from electron recoils (such as when irradiated by $\gamma$-rays). At higher energies (MeV and above), the scintillation output is generally found to be proportional to energy deposition but, at lower energies, a strong departure from proportionality has been observed for nuclear recoils. A thorough characterisation and understanding of such effects is essential for accurate low energy calibration, especially given the widespread use of scintillators in contemporary science. One specific example, where the low energy response to nuclear recoils is paramount,  can be found in the field of direct dark matter searches, both for the response of the dark matter targets themselves (e.g.~noble liquid scintillators), and where scintillators find their application in anti-coincidence detector systems~\cite{Akimov:2010p2728,Armengaud:2010p3115,Kozlov:2010p3116,Alner:2007}. It is in this context that the present results have been obtained. Conceptual designs for future, large active neutron rejection systems featuring scintillators are under discussion~\cite{Wright:2010}, and will require improved knowledge of the low energy response, even when the main neutron detection mechanism is via radiative capture. In the case of polystyrene-based scintillators, little data exist for recoils below $\sim$1~MeV, which are produced, for example, by radioactivity neutrons.

The present measurements were performed with the plastic scintillator used in the veto detector of the ZEPLIN--III dark matter experiment, based at the Boulby Mine, UK. ZEPLIN--III is a two-phase (gas/liquid) xenon detector designed to observe low-energy nuclear recoils from galactic weakly interacting massive particles (WIMPs)~\cite{Akimov:2007p58,Lebedenko:2009p156,Lebedenko:2009p157,Akimov:2010p180,Akimov:2011}. For its second science run the detector has been enclosed by a polystyrene, (C$_{8}$H$_{8}$)$_{n}$, based veto detector~\cite{Akimov:2010p2728}. The veto instrument includes 52 modules individually coupled to photomultiplier tubes (PMTs), totalling $\sim$1~tonne of scintillator, which, in the form of a barrel and a roof, surround the WIMP target. For a detailed discussion of the realised performance of the veto detector, the reader is referred to Ref.~\cite{Ghag:2011}.

\subsection{Quenching}

\noindent The scintillation light yield for a nuclear recoil of a given energy is quenched, {\em i.e.}, reduced in comparison to the scintillation output observed from an electron recoil of the same energy. A significant contribution to this difference may be identified with the heat associated with the atom cascades generated by nuclear recoils as described by Lindhard~\cite{Lindhard:1963}. A formalism in which the scintillation light yield of highly ionising particles depends not only on the energy of the particle but also on its stopping power in specific materials was developed by Birks~\cite{Birks:1951p2813}, of which a detailed description is presented in Ref.~\cite{Tretyak:2010p40-53}. The resulting relation may be written as:

\begin{equation}
\label{birks}
\frac{dL}{dr}=\frac{S\frac{dE}{dr}}{1+kB\frac{dE}{dr}} \,,
\end{equation}

\noindent where $dL/dr$ is the scintillation yield per unit path length $r$, $S$ is the absolute scintillation factor, $BdE/dr$ is the density of excitation centres along the recoil ionisation track and $k$ is a quenching factor. By finding the ratio between the light yield for electron recoils, $L_{e}$, and for ions, $L_{i}$, Eqn.~(\ref{birks}) may be rewritten in terms of the quenching factor for nuclear recoils, $Q_{i}$, in integrated form as:

\begin{equation}
\label{birks2}
Q_{i}=\frac{L_{i}(E)}{L_{e}(E)}=\frac{\int_{0}^{E} \frac{dE}{1+kB(\frac{dE}{dr})_{i}}}{\int_{0}^{E} \frac{dE}{1+kB(\frac{dE}{dr})_{e}}}.
\end{equation}

From (\ref{birks}) and (\ref{birks2}), an energy dependence of the quenching factor is apparent. This is especially significant for the low energy region where the stopping power experiences greatest variation. The majority of the measurements obtained to-date for the quenching factor in plastic scintillators concentrate on neutrons and protons in the energy region above $\sim$1~MeV~\cite{Rielly:1995,Smith:1969,Madey:1979,Knudson:1978,Miller:1968}.  In recent years, the need for precise knowledge of neutron quenching factors for materials used in the direct search for dark matter has led to significant new measurements at low energies, often making use of dedicated neutron scattering facilities~\cite{Simon:2003,Jagemann:2005}. However, no recent measurements have been reported for plastic scintillators despite their incorporation into several low energy event experiments. In this paper we  present measurements of nuclear recoil quenching factors for energies below 1~MeV down to 125~keV.

\section{Experimental set-up}

\noindent One of the 52 plastic scintillator modules of the \mbox{ZEPLIN--III} veto detector was used for data taking in the Boulby Underground Laboratory, an intrinsically low background environment. The scintillator bar has the form of a parallelepiped of length 1~m, width 15~cm, and a trapezoidal cross-section with parallel sides of length 15~cm and 12~cm. The polystyrene-based  scintillator material (UPS--923A, p-terphenyl 2$\%$, POPOP 0.02$\%$, produced by Amcrys-H, Kharkov, Ukraine \cite{Amcrys-Hwebsite}) has a density of 1.06~g/cm$^{3}$ and a refractive index of 1.52 for blue light. The average light output for electron recoils has been measured to be $\sim$5500~photons/MeV~\cite{Akimov:2010p2728}. The scintillation light shows a peak intensity at 420~nm, a rise time of 0.9~ns and a decay time of 3.3~ns. The average bulk attenuation length for the 52 modules has been experimentally measured and is found to be approximately 1~m~\cite{Akimov:2010p2728}.

To increase the effective attenuation length of the plastic and improve light collection, a specularly reflective aluminised Mylar foil is placed at one end. Additionally, the module has been wrapped in diffuse reflector PTFE sheet on all sides.  Light produced in the scintillator is detected with a PMT (ETEL--9302KB) of quantum efficiency 30$\%$~\cite{Akimov:2010p2728} optically coupled to the end opposite the mirror.

Energy spectra were recorded with the dedicated data acquisition system of the veto detector (CAEN model V1724), which digitises waveforms with 14-bit resolution, 0--2.25 V input range,  40~MHz bandwidth and a sampling rate of 10~MS/s. In this instance waveforms were parameterised using a bespoke data reduction software adapted from that developed for the ZEPLIN-III instrument~\cite{ze3ra}. The trigger was provided by an external pulse generator at a constant frequency. Additionally, during the neutron source measurements, data were taken simultaneously with a single-channel pulse height analyser (``MAESTRO SCA"), triggered by an internal discriminator.

To measure the response to nuclear recoils, the scintillator was exposed to neutrons from a $^{241}$Am-Be ($\alpha$,n) source and, separately, to a $^{252}$Cf fission source. The plastic was shielded from $\gamma$-ray emission from the sources and the environment by enclosing it in a 20~cm thick castle composed of low-background Cu and Pb in equal parts with an additional 4~cm of lead on the roof. Neutron exposures were performed with the sources placed directly on the castle ($\sim$50~cm above the sealed scintillator). Systematic uncertainties in the setup were explored extensively from which it was found that variation in neutron source position had negligible effect.
 
 Crucially, $\gamma$-ray attenuation and external electron-recoil contamination within the nuclear recoil data have been quantified using Monte Carlo simulations and dedicated measurements (see Sec.~III for detailed discussion of simulations). In particular, we examined the effect of varying the thickness of lead shielding placed over the castle. $\gamma$-ray emission spectra from the $^{252}$Cf and Am-Be sources (reconstructed from values given in the NuDat database \cite{nudat}) have been studied separately. The actual $\gamma$-ray activities were 21,000$\pm$2,100 $\gamma$/s for the $^{252}$Cf source and 6,300$\pm$400 $\gamma$/s for the Am-Be source (the latter accounts only for the two highest energy $\gamma$-rays of 3.21~MeV and 4.44~MeV from de-excitation of $^{12}$C$^{*}$ populated by the Be($\alpha$,n) reaction). The simulations indicate that a single $\gamma$-ray from the Am-Be source would be transmitted through the shielding along with every 30,000 neutrons (of which $\sim$600 deposited energy in the scintillator bar)  for the nominal lead thickness in our configuration, while no $\gamma$-rays from the $^{252}$Cf source exposure would be observed. Thus, the results show that the $\gamma$-ray fluxes from the sources make no significant contribution to the neutron exposure data. To confirm this conclusion, an extended exposure of the scintillator to a 11~kBq $^{60}$Co $\gamma$-ray source (1.17 and 1.33 MeV $\gamma$-rays) placed externally on the upper surface of the enclosure was performed.  No measurable increase in event rate over background was observed. Given that contributions from the $\gamma$-rays coming from the sources themselves are negligible in the neutron measurements, most $\gamma$-rays detected during the neutron exposure are generated internally (inelastic scattering and radiative neutron capture). Non source-related backgrounds, arising, for example, from low level activity of shielding components and the plastic scintillator itself, are measurable but not significant above a threshold of 2 photoelectrons (phe).

\section{Simulations}

\noindent The methodology used to extract the quenching factor was first applied to liquid argon scintillation by the WARP group \cite{Benetti:2008}; other examples followed \cite{Sorensen:2009p2801,Sorensen:2010,Lebedenko:2009p156,Horn:2011}. Experimental data are compared to a comprehensive Monte Carlo simulation that includes a detailed description of the experiment. The relationship between real energy deposition and resulting scintillation production, {\em i.e.}, the energy-dependent quenching factor, is included as a parameter in the simulation. An iterative process is used to optimise the quenching factor, minimising on $\chi^{2}$ in the comparison between data and simulated energy spectra. The simulations have been performed with the GEANT4 toolkit~\cite{geant4} (version 9.2, with neutron cross-sections from ENDF/B-VI~\cite{endf}) using standard neutron spectra for the two sources (Am-Be ISO 8529-1~\cite{ambeISO}, $^{252}$Cf fission spectrum from SOURCES-4C \cite{Wilson:1999}). Emitted neutrons and their secondaries are propagated including all relevant nuclear and electromagnetic physical processes; a set of optical processes describes the generation and detection of scintillation light from nuclear and electron recoil interactions in the scintillator. These photons are tracked to the photocathode of the PMT including relevant optical effects (reflection, refraction, attenuation) at which point the production of photoelectrons is simulated. Appropriate random fluctuations are included to model the production of scintillation photons and the production of photoelectrons from the PMT photocathode.

It should be noted that for a full description of neutron scattering in hydrogenous materials the standard GEANT4 elastic scattering process must be supplemented with a model (G4NeutronHPThermalScattering) to describe the energy region below 4~eV for the correct treatment of thermal neutron scattering from chemically-bound atoms. In these molecules several temperature-dependent  vibrational modes are possible, which alter the scattering cross-section~\cite{thermal}.  This is of particular relevance to this study and radiative capture on hydrogen is enhanced by $\sim$20\% over the standard model.

\section{$\gamma$-Ray Calibrations}

\noindent By definition, the response of the plastic scintillator to \mbox{$\gamma$-rays} is unquenched, allowing standard $\gamma$-ray sources to be used to determine the overall gain of the system. Moreover, it is expected that the GEANT4 simulations should provide an excellent match to the $\gamma$-ray calibration data, validating most processes included in the physics model and the accuracy of the geometry implementation. The PMT gain was set such that both single photoelectron (SPE) peaks and Compton edge features could be resolved in all spectra, allowing presentation of the data in terms of absolute numbers of photoelectrons.

\begin{figure}[h]
\begin{center}
\includegraphics[width=3.4in]{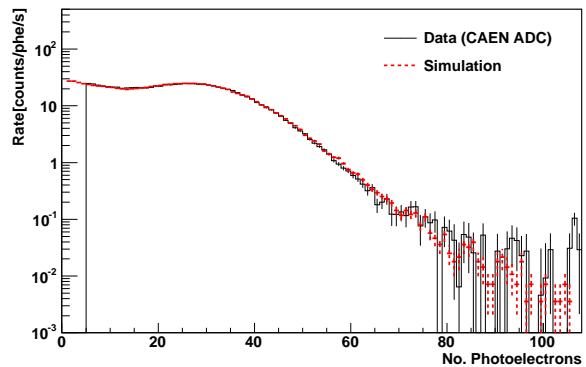}
\caption{Energy spectrum acquired from a $^{137}$Cs $\gamma$-ray source exposure. The data acquired with the CAEN acquisition (solid black spectrum) with a threshold of 5 photoelectrons is shown in comparison to the simulation data (red dashed spectrum).}
\label{cs137}
\end{center}
\end{figure}

With the roof of the shielding castle open, calibration measurements with a $^{137}$Cs $\gamma$-ray source (4.7~kBq) were performed.  Figure~\ref{cs137} shows the acquired spectra in comparison to Monte Carlo simulations. Data were acquired with the CAEN acquisition system (solid black spectra) with a trigger provided by an external pulse generator operating at constant frequency. Signal pulses were then extracted from the recorded waveforms. The result of a GEANT4 simulation of this exposure is shown by the dashed red line;  excellent agreement across the full energy range is demonstrated. The scintillator module used has an attenuation length of ~80~cm and the photoelectron yield with the calibration source above the centre of the plastic (48.7~cm from the photocathode face) is measured to be $\sim$44~phe/MeV.

The present simulations do not include spurious effects such as dark emission from the photocathode, after-pulsing, or $\beta^{-}$ radiation from $^{40}$K contamination in the glass PMT envelope. Each of these effects are known to contribute at the single to few photoelectron level with significant rate~\cite{Ghag:2011,Araujo:2011}. Consequently, we impose a 5 photoelectron analysis threshold on the $\gamma$-ray calibration data and, therefore, on our neutron scattering analysis and results.

\section{Neutron Exposures}

\subsection{Nuclear recoils}

\begin{figure}[htbp]
\begin{center}
\includegraphics[width=3.4in]{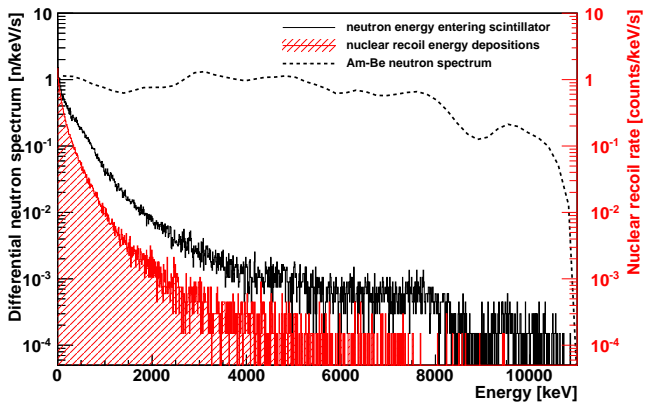}
\caption{Energy depositions in the scintillator from neutron-induced nuclear recoils coming from an Am-Be source (red hatched spectrum -- referring to the scale on the right). The y-axis on the left refers to the neutron flux from the source (black dashed spectrum) and the differential neutron spectrum when entering the scintillator bar (black solid spectrum).}
\label{ambespec}
\end{center}
\end{figure}

\begin{figure}[htbp]
\begin{center}
\includegraphics[width=3.4in]{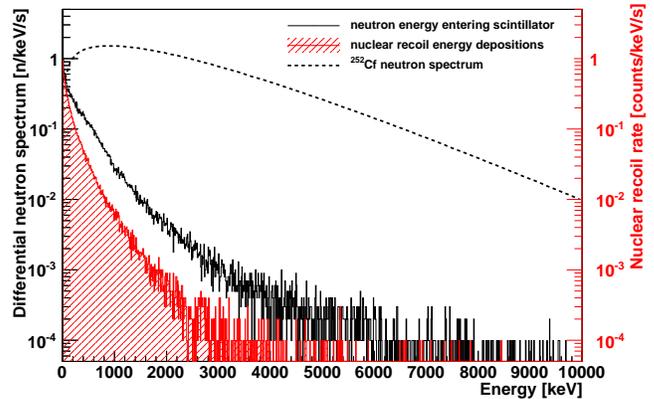}
\caption{Energy depositions in the scintillator from neutron-induced nuclear recoils coming from an $^{252}$Cf source (red hatched spectrum -- referring to the right hand scale). The left hand scale refers to the original neutron spectrum (black dashed spectrum) in comparison with the differential rate of neutrons entering the scintillator bar (black solid spectrum).}
\label{cf252spec}
\end{center}
\end{figure}

\noindent Data were accrued for a live time of 600~s from separate exposures to the Am-Be source (5,500$\pm$300~neutrons/s) and the $^{252}$Cf source (3,400$\pm$170~neutrons/s). Placing the sources externally to the copper-lead enclosure attenuates the $\gamma$-ray emission from the sources to a negligible level. The impact of the enclosure on the neutron fluxes is illustrated in Figs.~\ref{ambespec} and \ref{cf252spec} for the two sources. The figures show the neutron emission spectra, the energy spectra as they enter the scintillator (both referring to the y-axis on the left), and the resulting nuclear recoil energy depositions in the polystyrene (y-axis on the right). The spectra at the scintillator interface include single neutrons being recorded multiple times as they are scattered out of the scintillator and re-enter again after interacting with the shielding. The recorded energy depositions are the total integrated signal from each individual neutron-induced recoil event. The shielding attenuates significantly the neutron flux, and scattering reduces the energies of surviving particles. Since this is a large effect, we quantified how the uncertainty in the lead thickness affects the neutron spectrum at the scintillator interface.  Variations up to $\pm$0.5~cm are found to be statistically insignificant.

The impact of thresholds in the simulated neutron source spectra (50 keV in both instances) has been examined. Reasonable extrapolations down to 0 keV do not change the recoil spectrum above threshold much and the ensuing quenching factor analysis is affected very little.

\subsection{Quenching factor}

\begin{figure}[htbp]
\begin{center}
\includegraphics[width=3.4in]{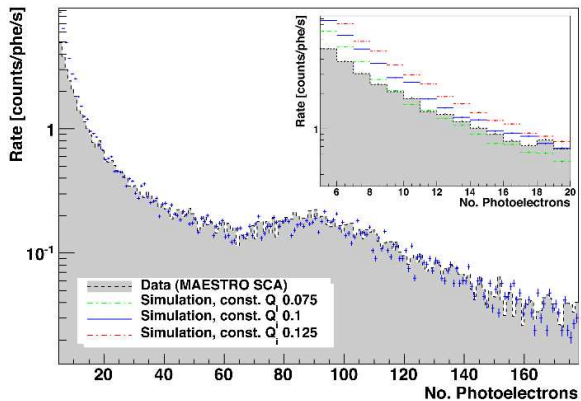}
\caption{Background-corrected energy spectrum originating from irradiation with an Am-Be source (grey shaded area) in comparison with simulations using the quenching factor $Q_{i}$ as a constant parameter for the whole energy range. The best agreement with the real data is met by the curve featuring $Q_i$=0.1 (blue solid spectrum). The peak at $\sim$90~phe is the 2.2~MeV radiative capture $\gamma$-ray from hydrogen. The inset shows the impact of different constant quenching factors at low photoelectron values. A marked discrepancy between simulation and data suggests that an energy-dependent quenching factor may provide a better physical description for low recoil energies.}
\label{ambeqf}
\end{center}
\end{figure}

\begin{figure}[htbp]
\begin{center}
\includegraphics[width=3.4in]{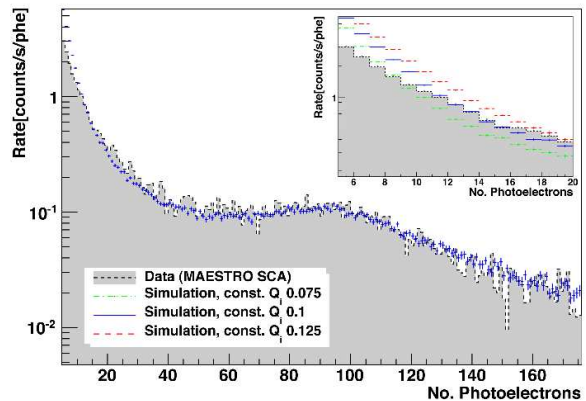}
\caption{Background-corrected energy spectrum originating from irradiation with a $^{252}$Cf source (grey shaded area) in comparison with simulations using the quenching factor $Q_{i}$ as a constant parameter for the whole energy range and a close up of the very low energy part of the spectrum as an inset at the top right.}
\label{cf252qf}
\end{center}
\end{figure}

\noindent 
Where data do not exist for quenching factors at low energies, it is customary to assume an energy-independent quenching as determined at higher energies. Various constant quenching factors have been considered and then compared to the present experimental data. Figures~\ref{ambeqf} and \ref{cf252qf} show the data from the Am-Be  and $^{252}$Cf source exposures in comparison to simulations which assumed an energy independent quenching factor, with $Q_i$=0.1 yielding, in both cases, the best fit to the measured data. For such a value, the nuclear recoil spectrum is quenched sufficiently such that the (un-quenched) peak at 2.218~MeV from  $\gamma$-ray emission following radiative capture of neutrons on hydrogen can be resolved.  This feature, appearing at $\sim$90~phe (with $\sigma$~$\sim$30~phe) may, thus, be used to normalise the energy scales of simulated to observed spectra and extract a quenching factor for the nuclear recoils. Both figures show that by adopting energy independent quenching factors, a discrepancy occurs below $\sim$35 phe. Above this value the goodness-of-fit is determined by statistical fluctuations only in both cases. The data shown in Figs.~\ref{ambeqf} and~\ref{cf252qf} were recorded with the MAESTRO SCA for the reason of better statistics at the position of the hydrogen capture peak. Subsequent analysis was mainly performed using data acquired with the CAEN system to avoid bias from threshold dependent trigger setups. Aside from counting statistics, the two recordings do not differ from each other at higher energies.

At very low photoelectron values ($\lesssim$20) greater divergence is observed between the the Monte Carlo and the measured data (see insets in Figs.~\ref{ambeqf} and \ref{cf252qf}) indicating an energy dependent behaviour of the quenching factor at low recoil energies. The methodology used to derive this energy dependent behaviour is as follows: a hypothetical $Q_{i}(E)$ function is composed from 14 values of recoil energy (125, 150, 175, 200, 225, 250, 300, 350, 400, 450, 550, 650, 750, 850 keV) and interpolated linearly between these points; a constant behaviour is assumed below and above this range. Above 1~MeV, low statistics and the decreasing gradient of the quenching factor preclude more in-depth analysis. For each combination of $Q_{i}(E)$ parameters (from a limited grid, guided to cover reasonable ranges), the full simulation is performed and $\chi^{2}$ calculated for the resulting match to the data. Below 5~phe, spontaneous SPE emission and other effects described in Sec.~IV can make a significant contribution to the experimental data and, therefore, this region is excluded from the minimisation. The $Q_{i}(E)$ parameters are modified for each iteration until no significant improvement in $\chi^{2}$ can be obtained. 

\begin{figure}[htbp]
\begin{center}
\includegraphics[width=3.4in]{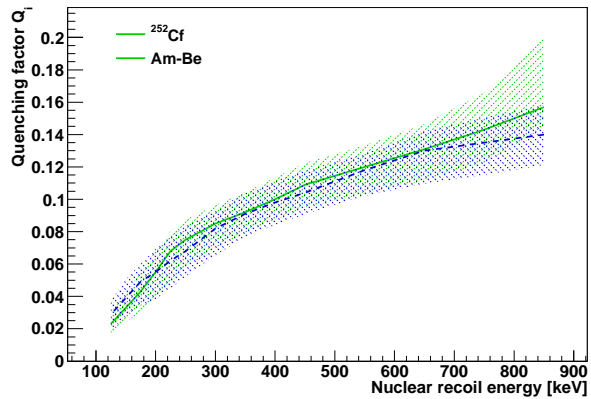}
\caption{The nuclear recoil quenching factor in polystyrene plastic scintillator UPS--923A as a function of recoil energy deposition extracted by mean of $\chi^{2}$ minimisation from comparison of simulations to data from a $^{252}$Cf (solid green) and an Am-Be (dashed blue) exposure, respectively. The hatched areas represent the 68$\%$ C.L. bands ($^{252}$Cf /, Am-Be $\setminus$ ).}
\label{qfedep}
\end{center}
\end{figure}

Figure~\ref{qfedep} shows the resulting energy-dependent quenching factor from minimising the overall $\chi^{2}$ for both datasets. Here the 5 phe analysis threshold allows measurements down to a nuclear recoil energy of approximately 125~keV. In the sub-threshold region below 5 phe an even stronger decrease in the quenching factor with energy would be inferred from uncorrected data. The 68$\%$ confidence intervals shown are determined by the envelope of regions built up from quenching factor model curves which fulfil the criterion of $\chi^{2}_{model}<\chi^{2}_{min}+Q_{y}$, where $Q_{y} $= 15.89 for 14 free parameters \cite{cowan}.

\begin{figure}[htbp]
\begin{center}
\includegraphics[width=3.4in]{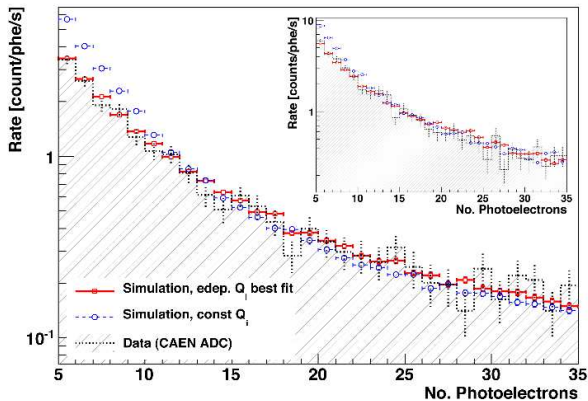}
\caption{Simulations using an energy-dependent value for $Q_{i}(E)$ in comparison with background corrected data acquired with the CAEN ADC (black hatched spectrum) from irradiation of the scintillator with a $^{252}$Cf and Am-Be source (inset), respectively. The best fit using $\chi^{2}$ minimisation is shown by the red solid histogram ($\Box$). For comparison, the blue dot-dash spectrum ($\bigcirc$) shows the use of a constant quenching factor (from best fit to data).}
\label{qfedep_cf252}
\end{center}
\end{figure}

Figure~\ref{qfedep_cf252} compares the  $^{252}$Cf data (black hatched histogram) with the best fit of the energy-dependent (red solid) and the best fit of the energy-independent simulation (blue dashed), respectively. The inset provides the same comparison but for the 
Am-Be study.

\subsection{Birks factor, $kB$}

\noindent Following the discussion in Ref.~\cite{Tretyak:2010p40-53}, the absolute value of the quenching factor for specific materials is expected to depend only on the so-called `Birks factor', $kB$, independently of the particle type. Consequently, the relative scintillation yield curve may be estimated by incorporating the appropriate energy-dependent stopping power for the specific particle species. The $kB$ factor is then determined by fitting Eqn. (\ref{birks2}) to experimental data.

\begin{figure}[htbp]
\begin{center}
\includegraphics[width=3.4in]{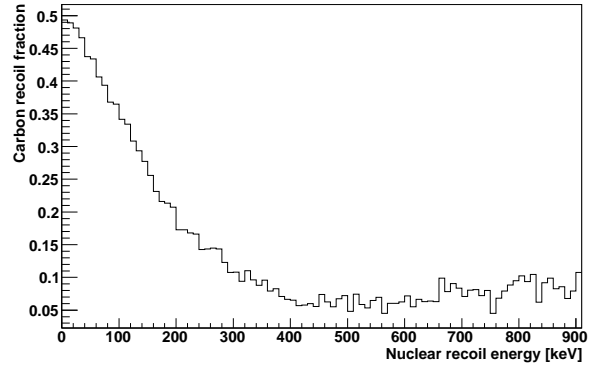}
\caption{Fraction of nuclear recoil energy depositions coming from carbon nuclei relative to the proton recoil contributions in the plastic scintillator averaged from exposures to both Am-Be and $^{252}$Cf neutron sources.}
\label{carbonfrac}
\end{center}
\end{figure}

At higher energies, contributions to the observed energy depositions come predominately from the scattering of protons in the plastic scintillator. For lower energy depositions, it is found that carbon nuclei (99$\%$ $^{12}$C) contribute over 30$\%$ of the overall nuclear recoil energy depositions. This relative fraction rises almost linearly in the lower energy region reaching $\sim$50$\%$ below 20 keV, as shown in Fig.~\ref{carbonfrac}.

\begin{figure}[htbp]
\begin{center}
\includegraphics[width=3.4in]{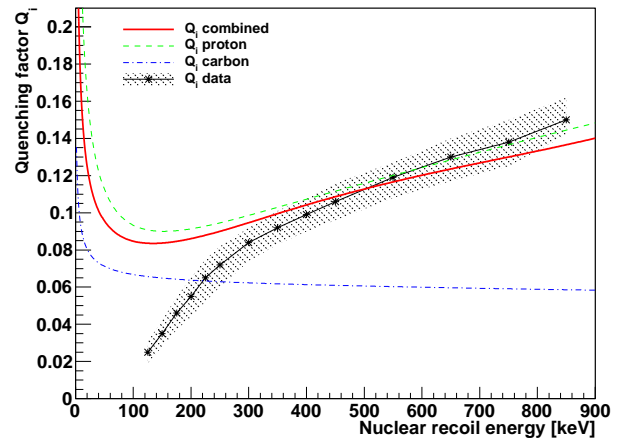}
\caption{A fit of semi-empirical calculations of Birks for combined proton and carbon stopping powers from varying the $kB$ factor (solid red) to the measured quenching factors (black with hatched error band) above 300 keV nuclear recoil energy yields :  $kB=0.0135$~g$\,$MeV$^{-1}$cm$^{-2}$. Additionally, curves assuming scattering off protons or off carbon nuclei only are also shown. Below this energy a clear divergence of measurement from the Birks description can be observed.}
\label{kbfit}
\end{center}
\end{figure}

The energy-dependent quenching factors derived here from the two neutron sources are in good agreement with each other and may therefore be combined. This is significant, since the neutron spectrum from an \mbox{Am-Be} source is somewhat uncertain below a few hundred keV, although this is especially so for stronger sources than the one used here~\cite{Marsh:1995}. The $^{252}$Cf fission spectrum, which is known more precisely, yields very similar results. A combination of the two results, following the prescription for asymmetric errors in Ref.~\cite{barlow:2003}, is presented in Fig.~\ref{kbfit} as the black solid line, with uncertainty represented by the shaded band. The quenching factor is seen to have a significant energy dependence, increasing in gradient towards low energies. In general, the observed dependence is reasonably similar to that expected from the Birks formalism above about 300~keV, but it departs from the expected behaviour at lower energies. Fitting the present results in the range of 300~keV to 850~keV results in a $kB$ factor of (0.014$\pm$0.002)~g$\,$MeV$^{-1}$cm$^{-2}$. The error given is statistical only. This is also shown in Fig.~\ref{kbfit}, with the contributions from protons, from carbon ions and the sum shown separately. Stopping powers for protons and carbon have been taken from NIST \cite{nist} and the SRIM Stopping Range Tables~\cite{SRIM}, respectively. 

The $kB$ factor resulting from fitting the present data to the Birks formalism above 300~keV may be compared with a previous value of $kB$=0.009~g$\,$MeV$^{-1}$cm$^{-2}$ reported for  $\alpha$-particle interactions in polystyrene-based plastic scintillator (\cite{Tretyak:2010p40-53} and references therein). The level of agreement is good, considering choice of data acquisition alone can produce discrepancies of a factor of two~\cite{Tretyak:2010p40-53}. The current results exhibit a slightly steeper dependence than expected from the Birks formalism, but interestingly, the same feature is apparent in all previous measurement for organic scintillators presented in the above reference. Above about 300~keV, the present data broadly support the semi-empirical description of Birks.

Below that nuclear recoil energy, a clear deviation from Birks is evident in Fig.~\ref{kbfit}, indicating that the fraction of scintillation generated by low energy nuclear recoils appears to decrease even more rapidly. As mentioned above, the analysis reported here has been limited to above 5~phe, to avoid complications that might be introduced by single photoelectron level processes not included in the simulations. However, not only would inclusion of these effects increase the discrepancy further, but examination of the 3--5 phe region indicates the trend continuing, with an even stronger dependence. A physical mechanism responsible for this behaviour is unclear. This is the first measurement to report on quenching factors at these energies for polystyrene.

\section{Conclusion}

\noindent
Despite the common use of plastic scintillators in industrial and scientific applications, little experimental data exist for the correlation between nuclear recoil energy deposition and scintillation output, especially below energies of 1~MeV.  Consequently, where plastic scintillators are used in low energy applications, a constant quenching of nuclear recoils is often assumed. We have measured the energy-dependent quenching factor for nuclear recoils in a polystyrene-based plastic scintillator \mbox{(UPS--923A)} for recoil energies between 125~keV and 850~keV. The analysis is based on comparison of observations of nuclear recoil spectra obtained with broad-band neutron sources with Monte Carlo simulations using the GEANT4 toolkit. Critical to this methodology is the accuracy of the Monte Carlo simulations; these demonstrated excellent reproduction of a $\gamma$-ray calibration source down to the analysis threshold of 5 photoelectrons. Significantly, the energy-dependent quenching factor for nuclear recoils was determined from measurements made with two different neutron source spectra, yielding the same result.  

We find that the Birks model describes reasonably the relation between energy deposition and non-radiative transfer processes over part of the energy range studied. A Birks factor $kB$=(0.014$\pm$0.002)~g$\,$MeV$^{-1}$cm$^{-2}$ was extracted from the best fit between semi-empirical calculations from a combination of proton and carbon nuclear recoils and the quenching factor curves presented. At lower energies a significant discrepancy between the Birks model and the present results was observed.\newline

\begin{acknowledgments}
\noindent The UK groups acknowledge the support of the Science \& Technology Facilities Council (STFC) for the ZEPLIN--III project and for maintenance and operation of the underground Palmer laboratory which is hosted by Cleveland Potash Ltd (CPL) at Boulby Mine, near Whitby on the North-East coast of England.  
The project would not be possible without the co-operation of the management and staff of CPL. Additionally, we want to thank the Boulby science facility team for their support during underground aspects of this work. We also acknowledge support from a Joint International Project award, held at ITEP and Imperial College, from the Russian Foundation of Basic Research (08-02-91851 KO\_a) and the Royal Society. LIP--Coimbra acknowledges financial support from Funda\c c\~ao para a Ci\^encia e Tecnologia (FCT) through the project-grants CERN/FP/109320/2009 and /116374/2010, and postdoctoral grants SFRH/BPD/27054/2006, /47320/2008, /63096/2009 and /73676/2010.
This work was supported in part by SC Rosatom, contract contract $\#$H.4e.45.90.11.1059 from 10.03.2011. The University of Edinburgh is a charitable body, registered in Scotland, with registration number SC005336.

\end{acknowledgments}

\end{document}